\documentclass[conference]{IEEEtran}
\usepackage[final]{graphicx}
\graphicspath{{Images//}} 

\usepackage{dirtree}
\usepackage{psfrag}
\usepackage{epstopdf}
\usepackage{amsfonts}
\usepackage{mathrsfs}
\usepackage{pifont}
\usepackage{verbatim}
\usepackage{upgreek}
\usepackage{color}
\usepackage[usenames,dvipsnames]{xcolor}
\usepackage{caption}
\usepackage{algorithmic}
\usepackage{algorithm}
\usepackage{mdwmath}
\usepackage{mdwtab}
\usepackage{amssymb,amsmath}
\usepackage{amsmath}
\usepackage{amsthm}
\usepackage{epstopdf}
\usepackage{cite}
\usepackage{scalefnt}
\usepackage{subfigure}
\usepackage{MnSymbol}
\usepackage{setspace}
\usepackage{mathtools}
\usepackage{psfrag}

\label{newcommand}

\label{English Chars}
\newcommand{\bA}{\textbf{A}}

\newcommand{\bD}{\textbf{D}}

\newcommand{\bF}{\textbf{F}}

\newcommand{\bG}{\textbf{G}}

\newcommand{\bP}{\textbf{P}}

\newcommand{\bQ}{\textbf{Q}}

\label{Roman Chars}

\newcommand{\bgamma}{\mbox{\boldmath{$\gamma$}}}

\newcommand{\bGamma}{\mbox{\boldmath{$\Gamma$}}}

\label{Theorems}

\theoremstyle{remark}

\begin{document}
\label{title}
\title{Distributed Resource Allocation Algorithms for Multi-Operator Cognitive Communication Systems}
\author{%
	\IEEEauthorblockN{Ehsan Tohidi\IEEEauthorrefmark{1}, 
		David Gesbert\IEEEauthorrefmark{1}, Philippe Ciblat \IEEEauthorrefmark{2}}\\
	\IEEEauthorblockA{\IEEEauthorrefmark{1}
		Communication Systems Department, EURECOM,
		06410 Biot, France,
		\{tohidi,gesbert\}@eurecom.fr}
		\IEEEauthorblockA{\IEEEauthorrefmark{2}
        Communications and Electronics Department, Telecom ParisTech, Paris, France,
        philippe.ciblat@telecom-paris.fr}
}
\maketitle
\begin{abstract}
\boldmath
We address the problem of resource allocation (RA) in a cognitive radio (CR) communication system with multiple secondary operators sharing spectrum with an incumbent primary operator. The key challenge of the RA problem is the inter-operator coordination arising in the optimization problem so that the aggregated interference at the primary users (PUs) does not exceed the target threshold. While this problem is easily solvable if a centralized unit could access information of all secondary operators, it becomes challenging in a realistic scenario. In this paper, considering a satellite setting, we alleviate this problem by proposing two approaches to reduce the information exchange level among the secondary operators. In the first approach, we formulate an RA scheme based on a partial information sharing method which enables distributed optimization across secondary operators. In the second approach, instead of exchanging secondary users (SUs) information, the operators only exchange their contributions of the interference-level and RA is performed locally across secondary operators. These two approaches, for the first time in this context, provide a trade-off between performance and level of inter-operator information exchange. Through the numerical simulations, we explain this trade-off and illustrate the penalty resulting from partial information exchange.
\end{abstract}

\section {Introduction}
Due to the increasing demand for higher data rates and scarcity of spectrum, CR communication systems have been of considerable interest during the last two decades \cite{1391031,7748543}. CR systems consist of a primary system that has the license of using the spectrum and also one or more secondary systems aiming to communicate in the same spectrum trying to maximize the data rate while guaranteeing not excessively interfering with the primary system. Quality of service (QoS) guarantee of the primary system is maintained through interference temperature thresholds which are held for each PU. Subsequently, the RA (subband assignment, power allocation, etc.) optimization problem is formulated with the SUs sum-rate as the objective function and a set of constraints to control the level of interference imposed on PUs. Although the optimization problem is usually NP-hard \cite{5963799}, considering a single operator scenario, a wide range of algorithms have been proposed to find a proper RA \cite{8546798,7336495,louchart2019resource}.


The RA problem becomes more challenging when several secondary operators are co-existing in an underlying manner with a common incumbent system, referred below as a \textit{multi-operator CR system}. In this case, SUs from different secondary operators contribute to the interference level at each PU and must coordinate to keep the total interference low. Generally speaking, two major approaches exist to tackle the multi-operator RA problem: centralized and distributed \cite{7748564}. In a centralized RA scheme, a central node in the network is responsible to manage the RA process. First, the central node receives SUs information of all operators and then, performs a centralized RA based on the given objective function and constraints. This approach has been significantly investigated in the literature \cite{7031944}. On the other hand, there is no center node for the distributed approach, and RA is performed in a decentralized manner that often cannot obtain the optimal solution. Several distributed approaches for RA in CR have been presented \cite{el2016distributed,gallego2012distributed}, where they either do not consider the coupling interference constraints or there is no flexibility in the level of information exchange among the operators. It should be pointed out that considering the coupling constraints, the information exchange among otherwise competing operators is often mandatory which
in most scenarios, raises complexity and privacy issues. 
Here, we investigate this problem in the particular context of cognitive satellite (CogSat) communications, where the secondary operators serve their subscribers via satellites, while the incumbent is a terrestrial cellular operator \cite{7060478}. 

In this paper, taking the challenge of inter-operator information exchange into account, we propose two approaches to tackle this problem. 
In the first approach, we propose an algorithm that shares a quantized version of  channel information to a central resource manager in order to split the interference level among the operators. Then, the resource allocation optimization is performed locally across the secondary operators based on the associated threshold of interference level. We demonstrate how the level of information shared in the first step affects the optimality of the achieved RA solution.
The second approach is to only share the level of imposed interference on PUs instead of exchanging SUs information. We propose two iterative algorithms in which the operators at each iteration share their contributed interference level and based on the interference threshold, a fusion center either allows them to increase their interference or force them to decrease it, and a new RA is performed at the operators locally. Although the central node in the second approach only performs a simple summation, both approaches need a central node. However, the main achievement of these approaches is a significant reduction in the level of shared data and for the first time in this context, enabling a trade-off between performance and level of shared data.
Through numerical results, we benchmark the performance of the proposed methods against the two extreme cases of information sharing: $(i)$ in comparison with the full information exchange algorithm (i.e., the centralized RA), and $(ii)$ a simple algorithm with no information exchange. We also demonstrate the tradeoff between SUs sum-rate and level of information exchange among operators.

\section{System Model}
Consider the problem of RA in a multi-operator CogSat communication system. We assume $L$ PUs forming the primary network and licensed to communicate over the spectrum. Moreover, $N$ secondary operators are considered each serving $K$ SUs and forming the secondary network. Each secondary operator has a satellite with $B$ beams sharing the same bandwidth. This bandwidth is split up into $M$ subbands. In each beam of each satellite, we assume a frequency division multiple access (FDMA) serving up to $M$ SUs, i.e., one subband per SU.

\begin{figure}
	\centering
	\includegraphics[width=.4\textwidth]{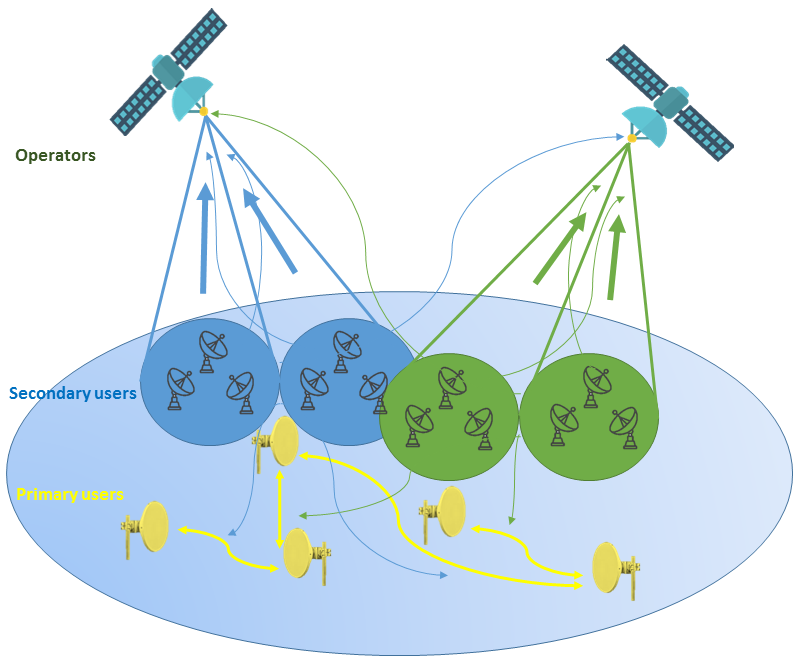}			
	\caption{A multi-operator CogSat communication system.}
	\label{fig:Scenario}
\end{figure}

An example of a CogSat communication system is depicted in Fig. \ref{fig:Scenario}, where $N=2$ operators each with $B=2$ beams are communicating with $M=3$ SUs per beam ($K=6$ SUs per operator). Also, there are $L=5$ PUs communicating through the incumbent network. In this paper, we consider the uplink channel for the secondary network.

We define $G_{k,b,m}$ as the channel gain in subband $m$ from SU $k$ to beam $b$ of the satellite serves SU $k$. Moreover, $F_{k,l,m}$ is the channel gain on subband $m$ from SU $k$ to PU $l$. The sets of all SUs, SUs of operator $n$, and SUs of beam $b$ of operator $n$ are denoted by $\mathcal{U}$, $\mathcal{U}_n$, and $\mathcal{U}_{n,b}$, respectively.


\section{Problem Formulation}
Considering the uplink channel, the goal is to maximize the sum-rate of SUs while not excessively interfering with the PUs. The optimization variables are subband assignment and power allocation which are as defined as follows:
\begin{itemize}
	\item $\bA \in\{0,1\}^{NK\times M}$ the subband assignment matrix where $A_{k,m}$ is $1$ if subband $m$ is assigned to the SU $k$ and $0$ otherwise.
	\item $\bP \in \mathbb{R}_+^{NK\times M}$ the power allocation matrix where $P_{k,m}$ corresponds to the transmission power of SU $k$ on subband $m$.
\end{itemize}
Therefore, the total interference imposed by the SUs on PU $l$ on subband $m$ is calculated in the following form
\begin{equation}
I_{l,m}=\sum_{k\in\mathcal{U}}{{A_{k,m}F_{k,l,m}P_{k,m}}}.
\label{interferPU}
\end{equation}

To guarantee the communication quality of the primary network, we consider interference-temperature constraints associated with the $L$ PUs and for each of the $M$ subbands. The constraints are represented as follows
\begin{equation}
I_{l,m} \leqslant \eta_{l,m}, \forall l,m,
\label{expectedinterference}
\end{equation}
where $\eta_{l,m}$ is the interference-temperature threshold at PU $l$ on subband $m$.

The signal power for SU $k$ of secondary operator $n$ in beam $b$, i.e., $k\in\mathcal{U}_{n,b}$, on subband $m$ at the satellite is $A_{k,m}G_{k,b,m}P_{k,m}$, while all transmissions by the other SUs at the satellite $n$ in beam $b$ play the role of interference for this user. Thus, for SU $k\in\mathcal{U}_{n,b}$, the received interference on subband $m$ for beam $b$ of satellite $n$ is given by
\begin{equation}
J_{n,k,b,m} = \sum_{i\in\mathcal{U}_n\setminus\{k\}}{A_{i,m} G_{i,b,m} P_{i,m}},
\label{interbeaminterference}
\end{equation}
where inter-operator interference is not considered since satellites are assumed to be far from each other and the SUs have highly directed radiation patterns toward their associated satellite \cite{maral2011satellite,7336495}. Furthermore, due to the FDMA scheme, there is no intra-beam interference among SUs.

We consider the sum-rate of SUs as the objective function.
The sum-rate for the secondary operator $n$ is calculated as follows
\begin{equation}
\begin{aligned}
R_n(\bG_n,\bA_n,\bP_n) =& \sum_{b=1}^B\sum_{k\in\mathcal{U}_{n,b}}{\sum_{m=1}^M{A_{k,m}}}\\\times&\log_2(1+\frac{G_{k,b,m}P_{k,m}}{1+J_{n,k,b,m}}),
\end{aligned}
\end{equation}
where $R_n(\cdot,\cdot,\cdot)$ is the sum-rate for operator $n$ and $\bG_n$, $\bA_n$, and $\bP_n$ are the channel gains, subband assignments, and power allocations of all SUs $k\in\mathcal{U}_n$, respectively.
Subsequently, the total sum-rate is given by
\begin{equation}
R = \sum_{n=1}^NR_n(\bG_n,\bA_n,\bP_n).
\end{equation}

Considering a peak power constraint on each subband for each SU, i.e., $0 \leqslant P_{k,m} \leqslant P_{\max}, \forall k,m$, the RA optimization problem can be formulated in the following form
\begin{equation}
\begin{aligned}
\max_{\bA,\bP}\quad & R \\
\text{s.t.}\quad& \text{C1: } \sum_{k\in\mathcal{U}}{{A_{k,m}F_{k,l,m}P_{k,m}}} 
\leqslant \eta_{l,m}, \quad \forall l,m, \\
& \text{C2: } A_{k,m} \in \{0,1\}, \quad \forall k,m, \\
& \text{C3: } 0 \leqslant P_{k,m} \leqslant P_{\max}, \quad \forall k,m, \\
& \text{C4: } \sum_{k\in\mathcal{U}_{n,b}}{A_{k,m}}=1, \quad \forall n,b,m, \\
& \text{C5: } \sum_{m=1}^M{A_{k,m}}=1, \quad \forall k, \\
\end{aligned}
\label{equ:mainopt}
\end{equation}
where C1 is to ensure that the interference level does not exceed the given threshold, C2 states that the subband assignment is binary, C3 is to limit the SU power between 0 and the maximum allowed level $P_{\max}$, and subband assignment restrictions, i.e., one SU in each beam be assigned to each subband and one subband be assigned to each SU, are applied in C4 and C5. Although the optimization problem \eqref{equ:mainopt} is known to be NP-hard \cite{5963799}, the main challenge is the need for information exchange among operators due to the set of coupling constraints C1. In fact, in order to find an optimal solution, operators have to share $\bG_n$ and $\bF_n$ with a central node. In this paper, we propose three algorithms to efficiently manage the exchanged information. 

\section{Proposed Algorithms}
In this section, we propose two different approaches that involve partial information exchanging among the operators.

\subsection{Channel Information Sharing}
Due to the objective function and constraint C1 in \eqref{equ:mainopt}, we need to have the channel gain matrices of all operators, i.e., $\bG$ and $\bF$, in a central node and apply a centralized algorithm to obtain $\bA$ and $\bP$. However, information sharing among operators causes a significant communication load and privacy issues. Therefore, we propose a partial information exchange scheme. More precisely, if $\gamma$ is an original form of data, we denote its quantized version by $\gamma^{(q)}$, where $q$ is the number of quantization bits (in case of a vector $\bgamma$ or a matrix $\bGamma$, $q$ quantization bits are used for each entry). Sharing the quantized version of information, the general idea is to first perform a centralized RA at the central node. Next, based on the centralized RA, split the interference-temperature thresholds among the operators. Then, since the coupling constraints are uncoupled, we perform a distributed RA across operators with their original information. In the following, the approach is explained in more detail.

Exchanging the quantized information, the centralized optimization problem is given as follows:
\begin{equation}
\begin{aligned}
\max_{\bA,\bP}\quad & \sum_{n=1}^NR_n(\bG^{(q)}_n,\bA_n,\bP_n) \\
\text{s.t.}\quad& \text{C1: } \sum_{k\in\mathcal{U}}{{A_{k,m}F^{(q)}_{k,l,m}P_{k,m}}} 
\leqslant \eta_{l,m}, \quad \forall l,m, \\
& \text{C2, C3, C4, C5,}
\end{aligned}
\label{equ:centralizedquantized}
\end{equation}
where C2-C5 are the same as in \eqref{equ:mainopt}.

Assume $\bA^{(q)}$ and $\bP^{(q)}$ are obtained by solving \eqref{equ:centralizedquantized}. Thus, based on $\bA^{(q)}$ and $\bP^{(q)}$, for all $l$ and $m$, each operator has a contribution to the interference level. In other words, $\eta_{l,m}$ is split up among SUs of operators. Consequently, share of each operator from interference-temperature thresholds is given below:
\begin{equation}
\eta^n_{l,m}=\sum_{k\in\mathcal{U}_n}{{A^{(q)}_{k,m}F^{(q)}_{k,l,m}P^{(q)}_{k,m}}},
\label{interfershare}
\end{equation}
where $\eta^n_{l,m}$ is the share of operator $n$ from interference-temperature threshold of PU $l$ on subband $m$.

Since operators are uncoupled in constraint C1 using the allocation in \eqref{interfershare}, we can formulate $N$ local optimization problems, one for each operator using the original information:
\begin{equation}
\begin{aligned}
\max_{\bA_n,\bP_n}\quad & R_n(\bG_n,\bA_n,\bP_n) \\
\text{s.t.}\quad& \text{C1: } \sum_{k\in\mathcal{U}_n}{{A_{k,m}F_{k,l,m}P_{k,m}}} 
\leqslant \eta^n_{l,m}, \quad \forall l,m, \\
& \text{C2, C3, C4, C5.}
\end{aligned}
\label{equ:localprecise}
\end{equation}
Since C2-C5 in \eqref{equ:mainopt} are satisfied in \eqref{equ:localprecise} for all operators, and C1 in \eqref{equ:mainopt} is satisfied as $\sum_{n=1}^N\eta^n_{m,l}\leqslant \eta_{m,l}$,
concatenating solutions of $N$ optimization problems \eqref{equ:localprecise} provides a solution for the original problem in \eqref{equ:mainopt}.

\subsection{Interference Level Sharing}
Apart from the optimality of the RA, the main reason for information exchange is to ensure interference-temperature constraints are satisfied. The idea of the second approach is that instead of channel gains information, we exchange the level of interference-temperature that each operator contributes, which enables performing RA locally for each operator.
There is less privacy issue in this approach as operators do not share $\bG$ and $\bF$ which prevents revealing location information of SUs of an operator for the other operators.
To tackle the RA problem through the interference-level sharing approach, we propose two optimization algorithms: $(i)$ Alternating direction method of multipliers (ADMM), and $(ii)$ Iterative equal-split. In the following, the proposed algorithms are introduced.

\subsubsection{ADMM}
is a well-known iterative optimization algorithm that is well suited to distributed convex optimization \cite{boyd2011distributed}. Here, we need to adapt the problem formulation in \eqref{equ:mainopt} to the standard form of the ADMM. Since in the standard form of the ADMM, the coupling constraints are in an equality form, we turn the coupling constraints in \eqref{equ:mainopt} into equalities by introducing a slack variable $\bD\in\mathbb{R}^{L\times M}$. Thus, the optimization problem can be reformulated as follows:  

\begin{equation}
\begin{aligned}
\max_{\bA,\bP}\quad & \sum_{n=1}^NR_n(\bG_n,\bA_n,\bP_n) \\
\text{s.t.}\quad& \text{C1: } \sum_{n=1}^N I^n_{l,m}(\bA_n,\bP_n) + D_{l,m}
= \eta_{l,m}, \quad \forall l,m, \\
& \text{C2, C3, C4, C5,}\\
& \text{C6: } 0 \leqslant D_{l,m} \leqslant \eta_{l,m}, \quad \forall l,m,\\
\end{aligned}
\label{equ:mainADMM}
\end{equation}
where $I^n_{l,m}(\bA_n,\bP_n)$ is the interference imposed by the SUs of operator $n$ on PU $l$ on subband $m$, i.e.,
\begin{equation}
I^n_{l,m}(\bA_n,\bP_n)=\sum_{k\in\mathcal{U}_n}{{A_{k,m}F_{k,l,m}P_{k,m}}}.
\label{interferPUN}
\end{equation}

Therefore, following the algorithm in \cite{falsone2019tracking}, at iteration $t$, the algorithm consists of two steps: the local step in which operator $n$ computes a minimizer of the following optimization problem
\begin{equation}
\begin{aligned}
&\bA_n^t,\bP_n^t \in \underset{\bA_n,\bP_n}{\arg\max}~ \{R_n(\bG_n,\bA_n,\bP_n) \\&- \sum_{l=1}^L\sum_{m=1}^M \lambda_{l,m}^{t-1}I^n_{l,m}(\bA_n,\bP_n) -\frac{c}{2}\sum_{l=1}^L\sum_{m=1}^M \\&||I^n_{l,m}(\bA_n,\bP_n)-I^n_{l,m}(\bA_n^{t-1},\bP_n^{t-1})+Q_{l,m}^{t-1}||_2\} \\
&\text{s.t.}\quad \text{C2, C3, C4, C5,}\\
\end{aligned}
\label{equ:ADMMfirst}
\end{equation}
where $\bQ\in\mathbb{R}^{L\times M}$ is a parameter of ADMM that gradually enforces the equality constraints (see \cite{falsone2019tracking}), $||\cdot||_2$ denotes $\ell_2$ norm, $c>0$ is a constant penalty parameter, and the superscript $t$ determines the iteration number. Also, the local step for the slack variable is performed as:
\begin{equation}
\begin{aligned}
&\bD^t \in \underset{\bD}{\arg\min}~ \{ \sum_{l=1}^L\sum_{m=1}^M\lambda_{l,m}^{t-1}D_{l,m} \\&+\frac{c}{2}\sum_{l=1}^L\sum_{m=1}^M||D_{l,m}-D_{l,m}^{t-1}+Q_{l,m}^{t-1}||_2\} \\
&\text{s.t.}\quad \text{C2, C3, C4, C5.}\\
\end{aligned}
\label{equ:ADMMfirstslack}
\end{equation}
Then, the central step where all the operators send their interference contributions on the PUs to a central node and the following updates take place:
\begin{equation}
\begin{aligned}
Q_{l,m}^{t} &= \frac{1}{N}(\sum_{n=1}^NI_{l,m}^{n}(\bA_n^{t},\bP_n^{t}) + D_{l,m}^t - \eta_{l,m})\\
\lambda_{l,m}^{t} &= \lambda_{l,m}^{t-1} + cQ_{l,m}^{t}.
\end{aligned}
\end{equation}
In this paper, we set a certain number of iterations for the ADMM algorithm to terminate.

\subsubsection{Iterative Equal-Split} is an iterative version of the equal-split algorithm. Let us start by introducing the equal-split algorithm. The equal-split algorithm is a single-step algorithm that does not require information exchange among the secondary operators.
The method starts with splitting up the interference-temperature thresholds equally among the operators and formulating the optimization problems locally for all operators, i.e., \eqref{equ:localprecise} with $\eta^n_{l,m} = \frac{\eta_{l,m}}{N}$.
In the equal-split algorithm, all the thresholds are split equally among the operators, however, a threshold can be a bottleneck for one operator while it is not limiting for another, and vice versa. Thus, the idea is to iterate on threshold splitting and see whether re-splitting the remaining threshold results in a sum-rate improvement.
For the iterative equal-split algorithm, at each iteration, we perform the equal-split algorithm for the remaining amount of the allowed interference. 
The iterative equal-split algorithm is as follows: at iteration $t$, calculating contribution of operator $n$ on PU $l$ on subband $m$ as \eqref{interferPUN}, we will have
$\sum_{n=1}^NI^n_{l,m}(\bA_n^{t},\bP_n^{t}) \leqslant \eta_{l,m}$.
Thus, the remaining threshold on subband $m$ of PU $l$ is
$\eta_{l,m}^{\rm{rem}} = \eta_{l,m} - \sum_{n=1}^NI^n_{l,m}(\bA_n^{t},\bP_n^{t})$.
Consequently, the threshold of operator $n$ for the next iteration is 
$\eta^n_{l,m} = \frac{1}{N}\eta_{l,m}^{\rm{rem}} + I^n_{l,m}(\bA_n^{t},\bP_n^{t})$.
We repeat this procedure for a given number of iterations.

\subsection{Solving Optimization Problems}
Although any algorithm that solves \eqref{equ:mainopt}, \eqref{equ:centralizedquantized}, \eqref{equ:localprecise}, \eqref{equ:mainADMM}, \eqref{equ:ADMMfirst}, and \eqref{equ:ADMMfirstslack} is applicable for our proposed methods, we perform convex optimization where due to the non-convexity of the objective function and constraints, we need to apply some well-known convex relaxation techniques;
One reason of non-convexity of the problem is due to the interference in \eqref{interbeaminterference}. Using the trick of \cite{louchart2019resource}, in satellite communication, we can neglect the inter-beam interference.
Also, we relax the binary constraints $A_{k,m}\in\{0,1\}$ to box constraints $0\leqslant A_{k,m}\leqslant 1$ and employ a change of variable $X_{k,m}=A_{k,m}P_{k,m}$. Then, we employ CVX to solve the convex optimization problems (any off-the-shelf solvers can also be employed). Since the obtained $\bA$ is continuous, we need a rounding algorithm to obtain a Boolean solution \cite{8537943} (projecting to the feasible domain). Here, we used the projection technique presented in \cite{louchart2019resource}. Finally, fixing the subband scheduling equal to the projected scheduling, we solve the optimization problem for $\bP$ only.

The total number of exchanged bits among operators and the central node for the proposed channel information sharing algorithm is:
\begin{equation}
\label{equ:totalexchange1}
    \begin{aligned}
    n_{\rm{exchanged}} &= N \times q  \times (K\times L \times M + K \times B \times M + L\times M),
    \end{aligned}
\end{equation}
where communication loads of sharing matrices $\bF_n$ and $\bG_n$ of all operators in the first step and allocated interference thresholds $\eta^n_{l,m}$ in the second step are considered. Further, for the interference level sharing algorithms, assuming $n_{\rm{iter}}$ iterations, the total number of exchanged bits is:
\begin{equation}
\label{equ:totalexchange2}
    \begin{aligned}
    n_{\rm{exchanged}} &= N \times q \times n_{\rm{iter}} \times L\times M.
    \end{aligned}
\end{equation}

\begin{figure}
	\centering
	\includegraphics[width=.42\textwidth]{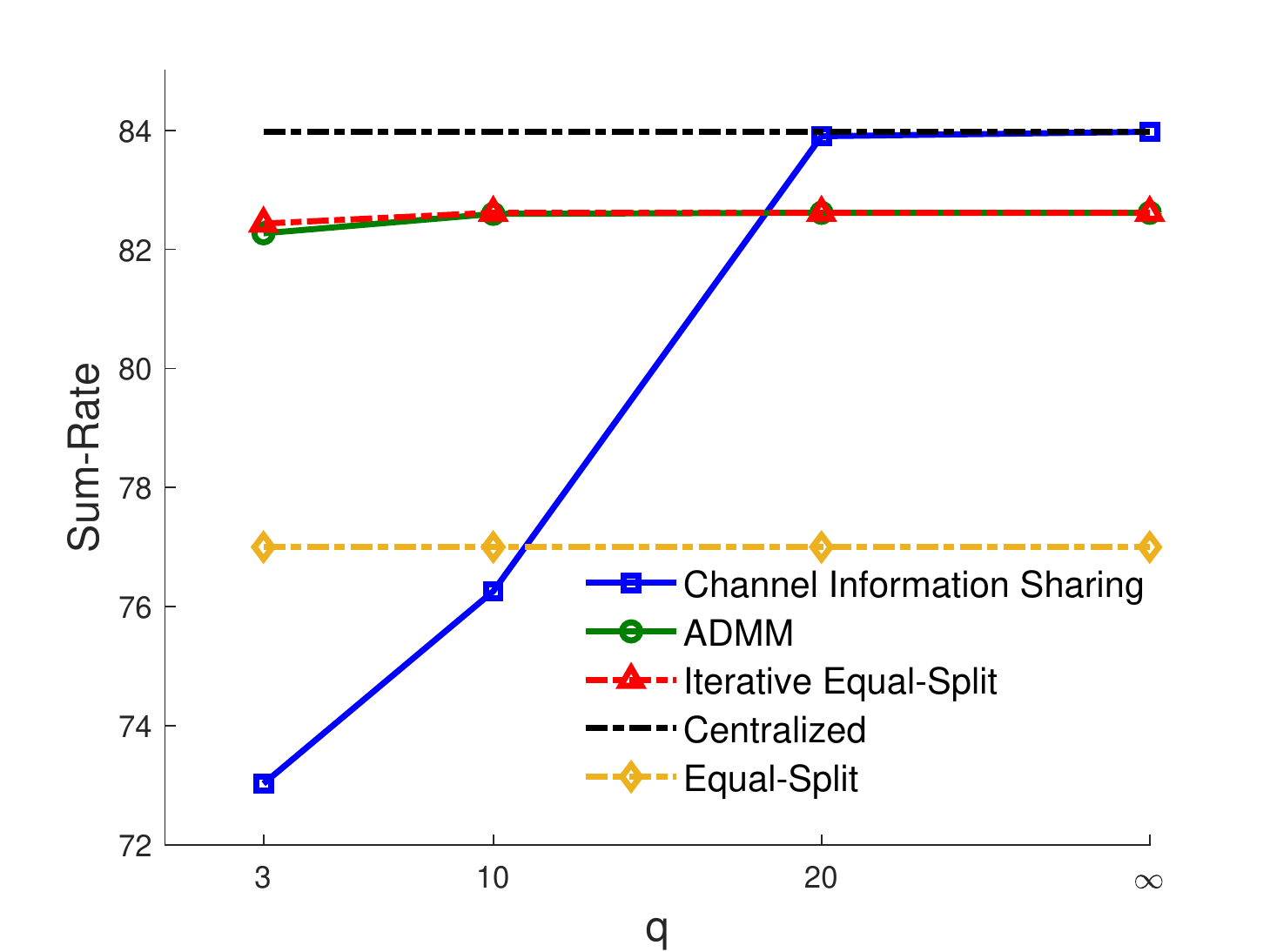}			
	\caption{Performance comparison of different algorithms; Sum-Rate versus the number of quantization bits $q$.}
	\label{SumRatCompVsQ}
\end{figure}

\begin{figure}
	\centering
	\includegraphics[width=.38\textwidth]{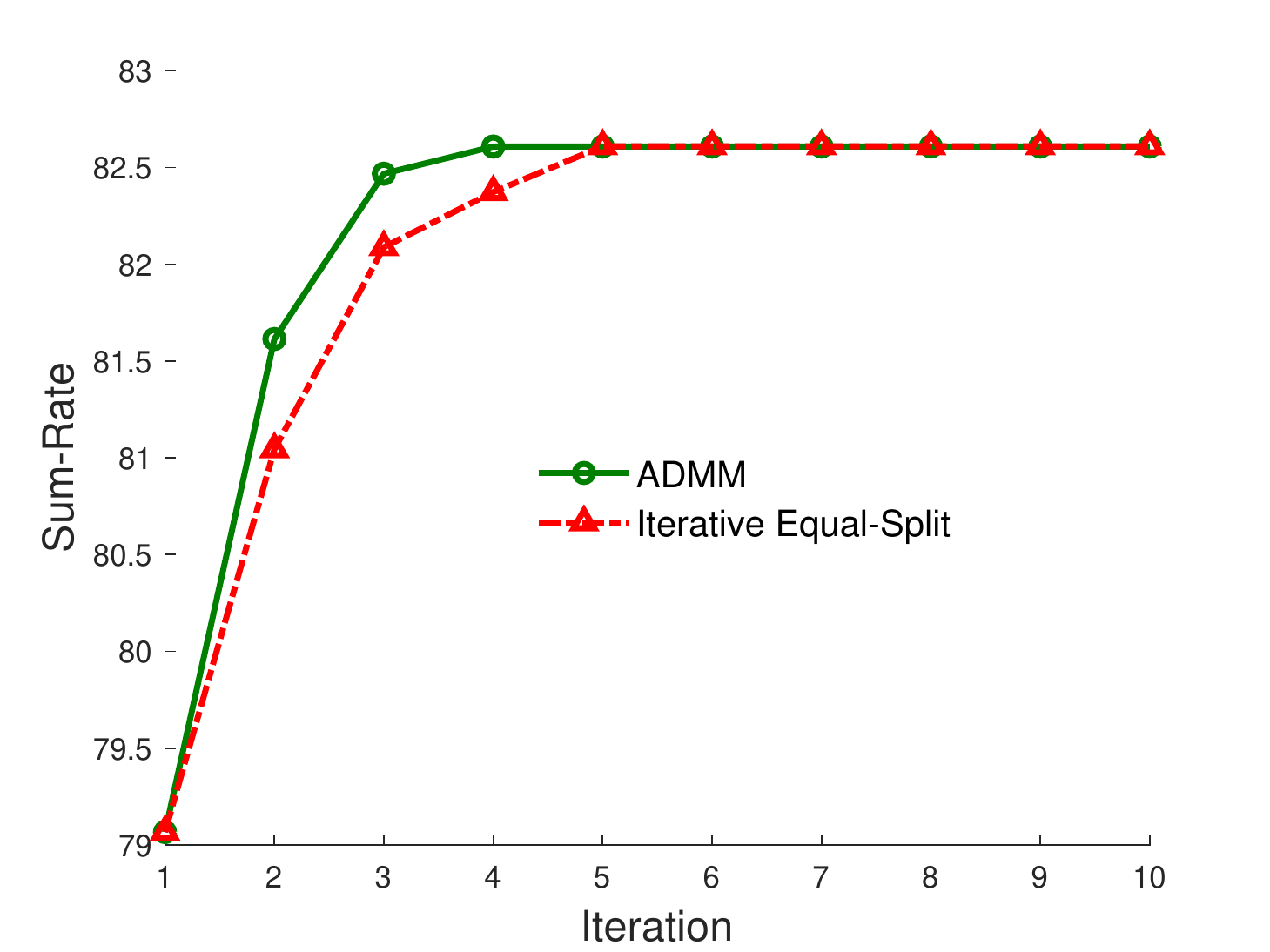}			
	\caption{Convergence rate of the interference threshold sharing algorithms.}
	\label{ConvergenceRate}
\end{figure}

\section{Numerical Results}
We consider a satellite system composed of $N=5$ operators, $B=2$ beams, $M=2$ subbands, $K=4$ SUs per operator, and $L=12$ PUs. The results of the proposed algorithms versus the number of quantization bits $q$ are plotted in Fig. \ref{SumRatCompVsQ} ($q\to\infty$ stands for an unquantized setting). 
All the proposed algorithms are quantized in such a way to make the total number of communicated bits equal (based on \eqref{equ:totalexchange1} and \eqref{equ:totalexchange2}). It should be pointed out that the $q$ presented in Fig. \ref{SumRatCompVsQ} is for the proposed channel information sharing algorithm which can be different from the corresponding $q$ for the interference level sharing algorithms. 
Since the simulation results show that the objective function is almost constant after iteration $5$ of the interference-level sharing algorithms (Fig. \ref{ConvergenceRate}), for comparison in Fig. \ref{SumRatCompVsQ}, we depict the result of the fifth iteration. Moreover, we present the result of centralized optimization and equal-split algorithms as the upper and lower bounds (i.e., extreme cases in terms of the level of the shared information) for the proposed algorithms, respectively. 
As expected, for the channel information exchange algorithm, increasing the number of quantization bits, the overall sum-rate increases, where with $q=20$, the result is almost equal to the centralized optimization case (i.e., $q\to\infty$). However, this improvement is marginal for the interference level sharing algorithms since the number of quantization bits is less critical for these algorithms.
On the other hand, exchanging interference-level of operators, sum-rate improvement is evident in comparison with the equal-split algorithm. Comparing the proposed algorithms, the main advantages of the channel information exchange algorithm is that exchanging information is done in two steps: $(i)$ in the first step when the quantized information of all the operators are shared in a fusion center, $(ii)$ and second, sending the result of centralized optimization for all operators. However, for the interference-level exchange algorithms, exchanging information among the operators occurs iteratively that can cause some delay in the procedure of RA. On the other hand, the main advantage of the interference-level exchange algorithms is not requiring a centralized optimization, and also, obtaining sum-rates close to the centralized solution. Moreover, a common advantage of both approaches is providing a trade-off that one can select the working point based on the demand (i.e., the desired sum-rate) and the imposed limitation (i.e., the communication capacity among operators which determines $q$).


\section{Conclusion}
In this paper, we have considered the challenge of inter-operator information exchange for CogSat RA and proposed two approaches to manage the level of information exchange. We demonstrated the existing trade-off between the level of information exchange and the sum-rate. An interesting observation is that it is possible to notably reduce the level of information exchange with a negligible performance loss.
\ifCLASSOPTIONcaptionsoff
\newpage
\fi
\bibliographystyle{IEEEbib.bst}
\bibliography{ref.bib}

\end{document}